# Performance improvement of three-body radiative diode driven by graphene surface plasmon polaritons

Ming-Jian He[1,2], Xue Guo[1,2], Hong Qi[1,2,*], Zhi-Heng Zheng[3,**], Mauro Antezza[4,5], and He-Ping Tan[1,2]

1 School of Energy Science and Engineering, Harbin Institute of Technology, Harbin 150001, P. R. China

2 Key Laboratory of Aerospace Thermophysics, Ministry of Industry and Information Technology, Harbin 150001, P. R. China

3 School of Energy and Materials, Shanghai Key Laboratory of Engineering Materials Application and Evaluation, Shanghai Polytechnic University, Shanghai 201209, China

4 Laboratoire Charles Coulomb (L2C), UMR 5221 CNRS-Université de Montpellier, F-34095 Montpellier, France

5 Institut Universitaire de France, 1 rue Descartes, F-75231 Paris, France

*Corresponding author: qihong@hit.edu.cn (H. Qi); zhiheng_zheng@163.com (ZH. Zheng)

**Abstract:** As an analogue to electrical diode, a radiative thermal diode allows radiation to transfer more efficiently in one direction than in the opposite direction by operating in a contactless mode. In this study, we demonstrated that, within the framework of three-body photon thermal tunneling, the rectification performance of three-body radiative diode can be greatly improved by bringing graphene into the system. The system is composed of three parallel slabs, with the hot and cold terminals of the diode coated with graphene films, and the intermediate body made of vanadium dioxide ($VO_2$). The rectification factor of the proposed radiative thermal diode reaches 300 % with a 350 nm separation distance between the hot and cold terminals of the diode. With the help of graphene, the rectification performance of the radiative thermal diode can be improved by over 11 times. By analyzing the spectral heat flux and energy transmission coefficients, it was found that the improved performance is primarily attributed to the surface plasmon polaritons (SPPs) of graphene. They excite the modes of insulating $VO_2$ in the forward-biased scenario by forming strongly coupled modes between graphene and $VO_2$, and thus dramatically enhance the heat flux. While, for the reverse-biased scenario, the $VO_2$ is at its metallic state and thus graphene SPPs cannot work by three-body photon thermal tunneling. Furthermore, the improvement was also investigated for different chemical potentials of graphene, and geometric parameters of the three-body system. Our findings demonstrate the feasibility of using thermal-photon-based logical circuits, creating radiation-based communication technology, and implementing thermal management approaches at the nanoscale.

## I. INTRODUCTION

For a long time, the physics of thermal radiation has been based on Planck's theory, which defines the maximum radiative heat flux between two blackbodies at different temperatures. However, in recent years, it has been found that in some extreme cases, when the separation between two objects is comparable to or even smaller than the peak wavelength of radiation, the classical theory is no longer applicable. In such cases, the heat flux can be much larger than the blackbody radiation limit[1-4]. This phenomenon is referred to as near-field radiative heat transfer (NFRHT)[5-9], in which evanescent waves take the place of propagating waves and dominate thermal radiation.

A thermal diode, which is the thermal analog of an electrical diode, is capable of rectifying the thermal current. In other words, the magnitude of the heat flux depends on the direction of the temperature gradient, such that the thermal diode transports heat flux mainly in one preferential direction rather than in the opposite direction[10, 11]. Based on the theory of NFRHT, radiative thermal diodes have been developed using different methods, such as utilizing doped silicon with different doping levels[12], the temperature dependence of the band edge absorption of indium antimonide[13], temperature-dependent coupled surface polariton modes[14], and asymmetric nanostructures[15]. Radiative thermal diodes not only have fundamental interest, but can also pave new ways for novel applications in thermal management and energy conversion at the nanoscale[16-18].

In recent years, phase transition materials have shown extraordinary advantages in modulating near-field radiative heat flux due to the non-linear dependence of physical properties on temperature[19, 20]. Furthermore, phase transition materials with the ability to exhibit different states have been utilized in the development of radiative thermal diodes[21-23]. Vanadium dioxide ($VO_2$), as a phase transition material, has been numerically demonstrated to be able to modulate near-field radiative heat flux by orders of magnitude upon rectification from the metallic to the insulating phase[24]. Inspired by this, a near-field thermal diode was developed using $VO_2$[25, 26], and its performance has also been experimentally verified[27]. However, as the separation distance increases, there is an obvious degradation in the performance of the radiative thermal diode. This degradation is mainly caused by the decline of NFRHT, in which the dominant modes can only be excited in the strong near-field regime and decay dramatically in the far field.

To overcome this limitation, we seek a solution from the emerging theory of three-body NFRHT, which exhibits more abundant and novel physical phenomena compared to the classic two-body system[28], due to the complexity of the heat transfer mechanism[29-31]. Messina and Antezza derived an expression for the near-field radiative heat flux of a three-body system in a thermal environment[32]. Based on this theory, an amplifier for photon heat tunneling was found based on a passive relay system[33]. Subsequently, utilizing the three-body theory and the photon thermal tunneling effect, micro/nano functional devices, such as thermal transistors[34-38], thermostats[39], thermal rectifiers[40-44], heat engines[45], and three-body near-field thermophotovoltaic systems[46, 47], were theoretically proposed. The photon thermal tunneling effect of the three-body system provides a new idea for improving the performance of the thermal diode by adding an intermediate body between the original two terminals to act as a relay.

Within the framework of three-body system, Gu at al. demonstrated that the functions of thermal switch and thermal rectification are able to be realized[41]. Latella et al. show that a strong asymmetry in the thermal conductance can appear in three-element radiative systems because of many-body interactions[44]. To further improve the performance of the three-body radiative diode, we take into account graphene, which is considered as the most representative two-dimensional material[48]. One of the most impressive properties of graphene is the surface plasmon polaritons (SPPs), which are localized surface waves generated by the resonance between electromagnetic waves and free electrons in graphene, that can be excited in the near-field regime[49, 50]. Moreover, the frequency of graphene SPPs overlaps with the peak wavelength of the electromagnetic wave emitted by the object at room temperature[51-54]. Based on this, graphene has shown superior performance in enhancing and modulating the near-field radiative heat flux by coupling SPPs with the modes of different materials[55-59]. Zheng et al. proved that by coating graphene on $VO_2$, which acts as a terminal in the two-body system, the rectification performance can be improved[23]. However, the graphene contacts with the $VO_2$ directly, and thus the optical properties of graphene may be affected during the phase transition.

In this paper, to improve the performance of the radiative diode and avoid direct contact between graphene and $VO_2$, we bring graphene to the three-body thermal diode, which consists of three parallel slabs, with the hot and cold terminals coated with a graphene film, and an intermediate body made of $VO_2$. We find that the performance of the radiative thermal diode can be greatly improved with the addition of graphene, due to the strong coupling between graphene SPPs and the modes excited by $VO_2$. Our paper is organized as follows: in section 2, we introduce the physical system and provide the critical formulas to simulate NFRHT. In section 3, we demonstrate the improved performance of the proposed thermal diode, analyze the physical mechanisms responsible for its operation, and investigate the influence of different parameters on its performance. Finally, in section 4, we provide concluding remarks.

## II. PHYSICAL SYSTEM

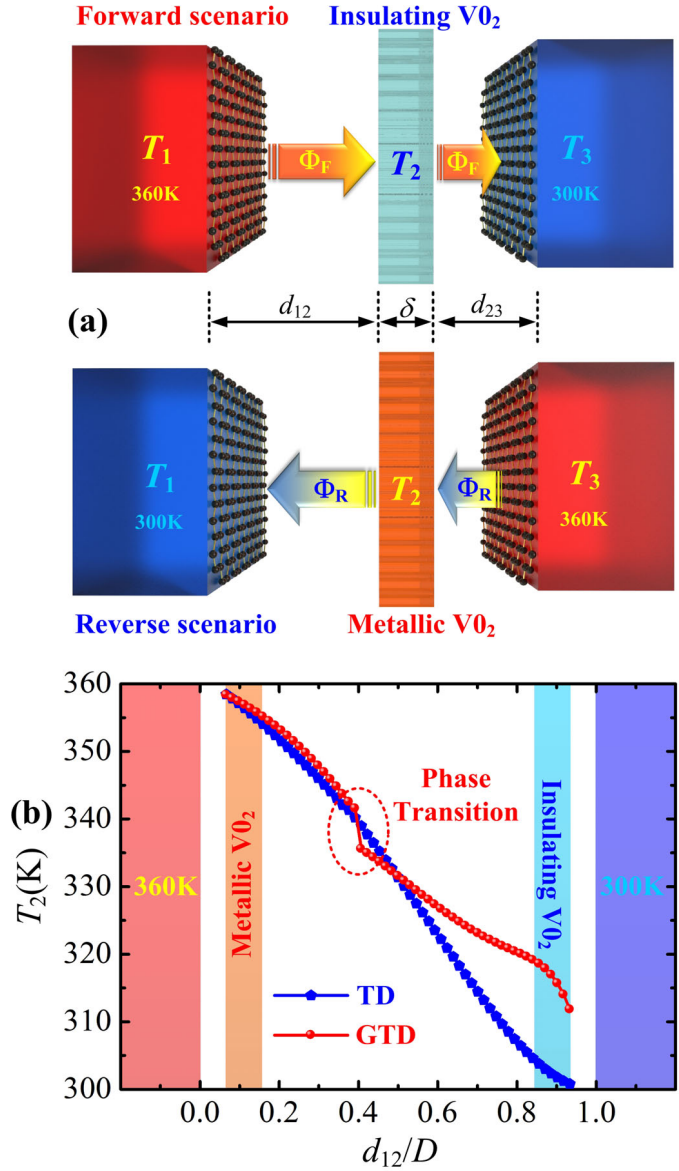

Fig. 1. (a) A schematic of a graphene-assisted thermal diode. The diode consists of three planar slabs at different temperatures: $T_1$ = 360 K, $T_3$ = 300 K, and the equilibrium temperature of the intermediate body $T_2$. The slabs on either side of the system are made of $SiO_2$ and covered with graphene, while the intermediate body is made of $VO_2$. The three slabs are separated by distances $d_{12}$, $d_{23}$, and the thickness of the intermediate body is denoted by $\delta$ = 50 nm. For the forward and reverse biased scenarios, $VO_2$ is in the insulating and metallic states, respectively, and the net radiative heat flux is denoted by $\Phi_F$ and $\Phi_R$. (b) The equilibrium temperature of the intermediate body $T_2$ at different positions of the system ($D = d_{12} + d_{23}$ = 300 nm) for a radiative thermal diode with and without graphene (with the chemical potentials of graphene set as 0.9 eV).

Here, we introduce the concept of a graphene-assisted thermal diode (GTD) in the framework of a three-body system. The schematic of the configuration is illustrated in Fig. 1(a) and is composed of three separated bodies labeled with indices 1-3. The three slabs are assumed to be infinite along the interfaces. The intermediate membrane with thickness $\delta$ is separated from the two external plates by separation distances $d_{12}$ and $d_{23}$, respectively. Hereinafter, if not specified, the total vacuum layer thickness $D = d_{12} + d_{23}$ is set as 300 nm, and the thickness of the intermediate body $\delta$ is 50 nm. It is challenging to implement such a thermal diode via three-body system, for the reason that the nanoscale vacuum layer and the parallelism of the three slabs are both hard to control based on the current nanotechnology. Nevertheless, the theoretical significance is bigger than practical significance for the work. The intermediate membrane is made of phase-change material VO$_2$, which undergoes an isulator-to-metal transition around its phase transition temperature, approximately equal to 341 K[60]. Below the phase transition temperature, VO$_2$ behaves like an insulator, with an anisotropic dielectric function given as a tensor[60]

$$\overline{\overline{\varepsilon}} = \begin{pmatrix} \varepsilon_\perp & 0 & 0 \\ 0 & \varepsilon_\perp & 0 \\ 0 & 0 & \varepsilon_\parallel \end{pmatrix} \quad (1)$$

where $\varepsilon_\perp$ and $\varepsilon_{//}$ are the dielectric function components vertical and parallel to the optic axis of uniaxial insulating VO$_2$, respectively. They can be modeled by the Lorentz model $\varepsilon(\omega) = \varepsilon_\infty + \sum_{j=1}^{N}\left[\left(S_j\omega_j^2\right)/\left(\omega_j^2 - i\gamma_j\omega - \omega^2\right)\right]$ [60]. Above the phase transition temperature, VO$_2$ behaves like a metal, with an isotropic dielectric function described by the Drude model[60]

$$\varepsilon(\omega) = -\left(\varepsilon_\infty \omega_p^2\right)/\left(\omega^2 + i\omega\omega_c\right) \quad (2)$$

The bodies on either side of the system are assumed to be semi-infinite plates made of silicon dioxide (SiO$_2$)[61] and covered with a graphene film. As for the graphene, it is modeled by a conductivity $\sigma_G$, as a sum of intraband and interband contributions, with respect to the angular frequency $\omega$, chemical potential $\mu$, and temperature $T$[49]

$$\sigma_G(\omega) = \sigma_{\text{intra}}(\omega) + \sigma_{\text{inter}}(\omega) \quad (3)$$

$$\sigma_{\text{intra}}(\omega) = \frac{2ie^2 k_B T}{(\omega + i\tau^{-1})\pi\hbar^2} \ln\left[2\cosh\left(\frac{\mu}{2k_B T}\right)\right] \quad (4)$$

$$\sigma_{\text{inter}}(\omega) = \frac{e^2}{4\hbar} \left[ G(\frac{\hbar\omega}{2}) + \frac{4i\hbar\omega}{\pi} \int_0^{+\infty} \frac{G(\eta) - G(\frac{\hbar\omega}{2})}{(\hbar\omega)^2 - 4\eta^2} d\eta \right] \tag{5}$$

where $G(x) = \sinh(x/k_B T)/[\cosh(\mu/k_B T) + \cosh(x/k_B T)]$, $e$ and $k_B$ are elementary charge and Boltzmann constant, respectively. $\hbar$ is Planck's constant divided by $2\pi$, and relaxation time $\tau$ is $10^{-13}$ s[49]. If not specified, the chemical potential of graphene is set to 0.9 eV hereafter, which can be modulated from ideally zero to 1eV by applying a bias voltage[48, 62].

Now let us consider the physics to simulate the radiative heat transfer in the proposed device. Based on the three-body theory[32, 33], the net radiative heat flow lost or received by body 3 can be can be expressed as

$$\Phi_3 = \int_0^\infty \frac{d\omega}{2\pi} \varphi_3(\omega) \tag{6}$$

with the spectral radiative heat flux given by[33, 63]

$$\varphi_3(\omega) = \hbar\omega \sum_{j=s,p} \int \frac{d^2\kappa}{(2\pi)^2} \left[ n_{12}(\omega) \xi_j^{1-2}(\omega,\kappa) + n_{13}(\omega) \xi_j^{1-3}(\omega,\kappa) \right] \tag{7}$$

In the expression listed above, $n_{ij}(\omega) = n_i(\omega) - n_j(\omega)$ represents the difference between the two mean photon occupation numbers $n_{i/j}(\omega) = (e^{\hbar\omega/k_B T_{i/j}} - 1)^{-1}$, with $i, j = 1, 2, 3$. $\xi_j^{\alpha-\beta}(\omega,\kappa)$ denotes the energy transmission coefficients between body $\alpha$ and $\beta$ at a parallel wave vector $\kappa$ and angular frequency $\omega$, in the polarization state $j$ ($j = s, p$). The energy transmission coefficients for both propagating waves $\kappa < \omega/c$ and evanescent waves $\kappa < \omega/c$ between different bodies can be calculated[32, 42]

$$\xi_j^{1-2}(\omega,\kappa) = \begin{cases} \dfrac{(1-|\rho_1^2|)(1-|\rho_2^2|)}{|D_{1,23}|^2}, & \kappa < \dfrac{\omega}{c} \\[2ex] \dfrac{4\text{Im}(\rho_1)\text{Im}(\rho_2)e^{-2\text{Im}(k_z)d_{12}}}{|D_{1,23}|^2}, & \kappa > \dfrac{\omega}{c} \end{cases} \tag{8}$$

$$\xi_j^{1-3}(\omega,\kappa) = \begin{cases} \dfrac{|\tau_2|^2 (1-|\rho_1|^2)(1-|\rho_3|^2)}{|D_{1,23}|^2 |D_{2,3}|^2}, & \kappa < \dfrac{\omega}{c} \\[2ex] \dfrac{4|\tau_2|^2 \text{Im}(\rho_1)\text{Im}(\rho_3)e^{-2\text{Im}(k_z)D}}{|D_{1,23}|^2 |D_{2,3}|^2}, & \kappa > \dfrac{\omega}{c} \end{cases} \tag{9}$$

where $D_{1,23}$ and $D_{2,3}$ are the Fabry-Pérot-type matrices expressed as follows[42]

$$D_{1,23} = 1 - \rho_1 \rho_{23} \exp[-2\text{Im}(k_z)d_{12}] \tag{10}$$

$$D_{2,3} = 1 - \rho_2 \rho_3 \exp[-2\mathrm{Im}(k_z)d_{23}] \tag{11}$$

$\rho_{23}$ is defined as the reflection coefficient of the unit consisting of bodies 2 and 3, as defined by $\rho_{23} = \rho_2 + |\tau_2|^2 \rho_3 \exp[-2\mathrm{Im}(k_z)d_{23}]/|D_{2,3}|^2$ [42]. Here, $k_z = \sqrt{\omega^2/c^2 - \kappa^2}$ is the $z$ component of the wave vector in vacuum. The reflection coefficient $\rho_\alpha$ and transmission coefficient $\tau_\alpha$ associated with a specific body are provided in detail in Ref. [33] and will not be repeated here. However, it should be mentioned that the formulas need to be modified for the insulating VO$_2$ due to its the uniaxial characteristics[23]. Additionally, the reflection coefficient of the external bodies $\rho_\alpha$ ($\alpha$ = 1, 3) needs to be modified due to the presence of the graphene sheet[51, 64]

$$\rho_{\alpha,s} = \frac{k_z - k_z^\alpha - \mu_0 \sigma_G \omega}{k_z + k_z^\alpha + \mu_0 \sigma_G \omega} \tag{12}$$

$$\rho_{\alpha,p} = \frac{\varepsilon_\alpha k_z - k_z^\alpha + \sigma_G k_z k_z^\alpha/(\varepsilon_0 \omega)}{\varepsilon_\alpha k_z + k_z^\alpha + \sigma_G k_z k_z^\alpha/(\varepsilon_0 \omega)} \tag{13}$$

Here, $k_z^\alpha = \sqrt{\varepsilon_\alpha \omega^2/c^2 - \kappa^2}$ is the normal component of the wave vector in medium $\alpha$ with permitivity $\varepsilon_\alpha$, where $\varepsilon_0$ is the vacuum permittivity, and $\mu_0$ is the vacuum permeability. In order to demonstrate the performance improvement caused by the graphene of GTD, the thermal diode without graphene (TD) is also investigated and compared to GTD in the following.

Similarly to $\Phi_3$, the net radiative heat flow lost or received by body 1 $\Phi_1$ can be obtained by changing the superscripts corresponding to different bodies in Eqs. (7)-(11)[33]. According to the three-body theory[33], the heat flow on intermediate body $\Phi_2$ can be expressed as the difference between $\Phi_1$ and $\Phi_3$, i.e., $\Phi_2 = \Phi_1 - \Phi_3$. In the steady state, the heat flux lost by the hot body equals that received by the cold one ($\Phi_1 = \Phi_3$), which means that the net heat flow dissipated in body 2 vanishes ($\Phi_2 = 0$). Then the intermediate body reaches its equilibrium temperature, labeled as $T_2$ in Fig. 1(a).

In Fig. 1(b), we assume that the hot and cold bodies are held at 360 K and 300 K (illustrated with the two colored squares), respectively, and are located on the left and right sides of the system. To realize the function of the proposed diode, the temperatures of the hot and cold bodies must be higher and lower than the phase transition temperature of VO$_2$. For simplicity, the effect of the operating temperature is not considered in the present work. The equilibrium temperature of the VO$_2$ membrane is plotted for different positions between the hot and cold bodies. The blue and red curves represent the equilibrium temperature of the VO$_2$ membrane for TD and GTD, respectively. The temperature of the intermediate body decreases as it moves from the hot body to the cold one. As a result, the VO$_2$ membrane converts from a metallic state to an insulating state when it is near the hot and cold terminals, respectively, which is demonstrated by the colored squares in Fig. 1(b). It should be noted that the phase transition is not instantaneous but takes place over a range of temperatures. The Bruggeman effective medium theory is used to model the effective dielectric constant of the VO$_2$ during the phase transition[65]. Although it is a method based on effective medium theory, it is different with the existing NFRHT works, which limit the separation distance much larger than the characteristic length of the microstructure. During the phase transition, the intermediate slab is viewed as a uniform mixture of insulating and metallic VO$_2$, and thus avoid the limitation of geometric parameter. An interesting phenomenon has been observed for GTD, where an obvious temperature step occurs near the phase transition temperature of VO$_2$. Moreover, the temperature curve of GTD has different changing trends when the VO$_2$ membrane is in the metallic and insulating states. In contrast, the $T_2$ curves of TD have basically the same changing trend in the two states and are smooth around the phase transition region. The different temperature responses of the VO$_2$ membrane can be attributed to the presence of the graphene. Compared to the results of TD, the equilibrium temperature of GTD is much higher, especially in the insulating state. This indicates that the heat transfer has been greatly enhanced when the graphene is added to the system.

## III. RESULTS AND DISCUSSION

### A. Performance and Physical Mechanism

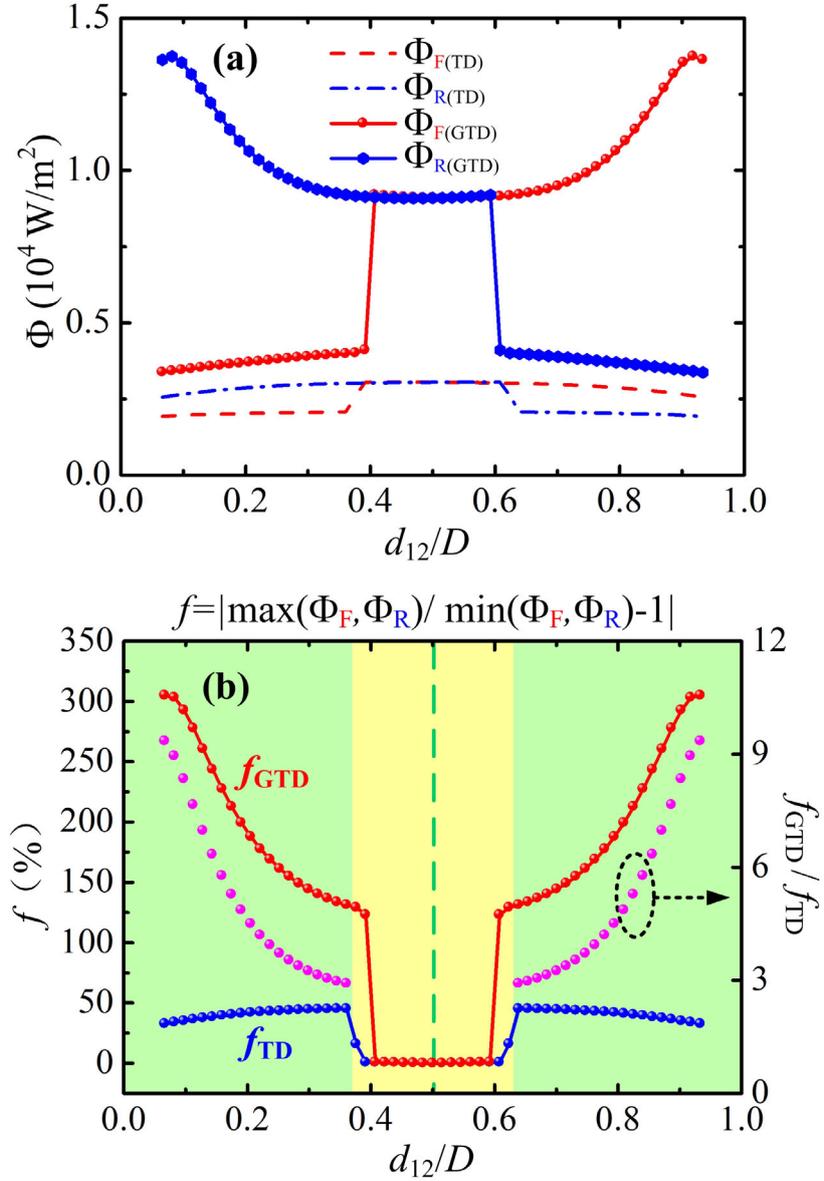

Fig. 2. When the VO$_2$ membrane is positioned at different locations within the three-body system, (a) the net radiative heat fluxes for systems with forward-biased (red) and reverse-biased (blue) temperature differences; (b) the rectification factors for TD and GTD, along with the ratio between them.

Now let us consider two different scenarios of the three-body system: (i) the forward-biased temperature difference scenario ($T_1$ = 360 K and $T_3$ = 300 K) and (ii) the reverse-biased temperature difference scenario ($T_1$ = 300 K and $T_3$ = 360 K), illustrated in Fig. 1(a). As introduced above, the system reaches the steady state when the heat flux lost by the hot body equals that received by the cold one ($\Phi_1 = \Phi_3$). Under this circumstance, the equilibrium radiative heat flow transporting in the system can be denoted as $\Phi_F$ and $\Phi_R$, with $\Phi_F = \Phi_1 = \Phi_3$ and $\Phi_R = \Phi_1 = \Phi_3$ for the two scenarios respectively. It should be stressed that the subscripts F and R in $\Phi_F$ and $\Phi_R$, denote the directions of temperature difference scenarios illustrated in Fig. 1(a), but not the forward-biased or reverse-biased direction of a diode. That is to say, the subscripts F and R have nothing to do with the magnitudes of the heat flow.

In Fig. 2(a), the net radiative heat flux ($\Phi_F$, $\Phi_R$) is plotted for the forward-biased (in red) and reverse-biased (in blue) temperature difference scenarios, when the $VO_2$ membrane is located at different positions in the three-body system. When $d_{12}$ is large and $d_{12}/D$ approximately equals to 1, just like the configuration in Fig. 1(a), the $VO_2$ membrane is very close to the right side of the system. In this situation, the temperature of the $VO_2$ membrane is close to the right-cold body (below the phase transition temperature of $VO_2$, $T_2$ < 341 K) in the forward-biased scenario and close to the right-hot body (above the phase transition temperature of $VO_2$, $T_2$ > 341 K) in the reverse-biased state. In other words, the $VO_2$ membrane is insulating and metallic in the forward-biased and reverse-biased scenarios, respectively. It can be seen from the results that the forward-biased heat flux $\Phi_F$ is much larger than the reverse-biased heat flux $\Phi_R$ in this situation, i.e., $\Phi_F > \Phi_R$. The curves have the same trend on either side of $d_{12}/D$ = 0.5, for the reason that the system is symmetric from the two different directions for the two scenarios. When the $VO_2$ membrane is located near the left side of the system, with $d_{12}/D$ approximate equal to zero, then the system exhibits an opposite style with the above, and $\Phi_F < \Phi_R$. No matter which is larger between $\Phi_F$ and $\Phi_R$, the heat flow transfers much more easily in one direction than in the opposite one, just like the current in the electrical diode. The phase transition of the $VO_2$ membrane results in two totally different heat transfer mechanisms, and hence the net radiative heat flux differs dramatically in the two scenarios. This is the basic principle of the radiative thermal diode proposed in the present work.

To compare the performance of TD and the proposed GTD, we present the results for the two different kinds of thermal diodes. The four curves in Fig. 2(a) demonstrate a significant enhancement of heat flux when graphene is used, regardless of whether the diode is forward-biased or reverse-biased. This implies that the introduction of graphene creates a new channel for energy exchange. Additionally, the difference between $\Phi_F$ and $\Phi_R$ is more pronounced when graphene is incorporated. To compare the performance between TD and GTD, we define the rectification factor as[13, 25, 26]

$$f = \left| \frac{\max(\Phi_F, \Phi_R) - \min(\Phi_F, \Phi_R)}{\min(\Phi_F, \Phi_R)} \right| \quad (14)$$

and present the results for different positions of the VO$_2$ membrane in Fig. 2(b). The red and blue lines indicate the thermal rectification factors for GTD and TD, respectively. The maximum values of $f_{GTD}$ and $f_{TD}$ are approximately to 300% and 50 %, respectively, indicating that the use of graphene significantly improves the performance of the radiative thermal diode. The ratio of $f_{GTD}$ to $f_{TD}$ is also plotted in Fig. 2(b), and the results show that the improvement due to graphene can be more than 9 times. Moreover, the $f_{GTD}$ and $f_{TD}$ curves exhibit different varying trends with respect to the positions of the VO$_2$ membrane, which provides further evidence that graphene alters the heat transfer mechanism in the system.

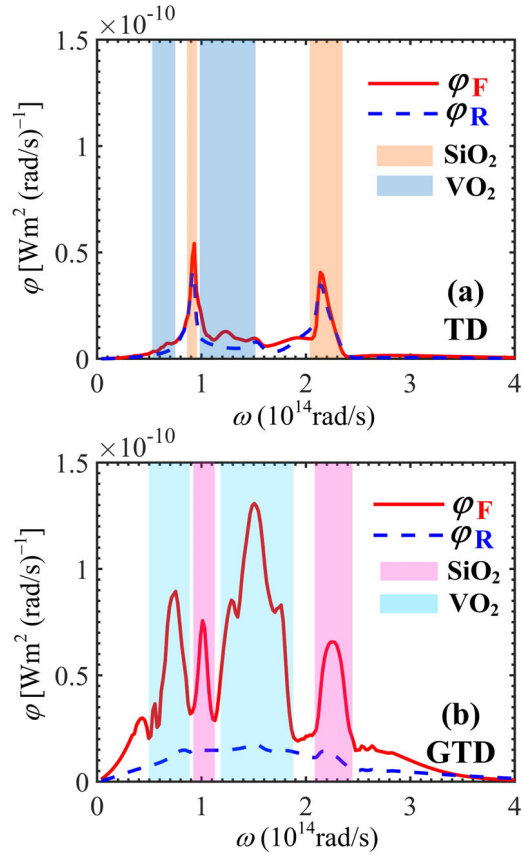

Fig. 3. The spectral radiative heat flux for (a) TD and (b) GTD, in forward-biased mode (solid lines) and reverse-biased mode (dashed line) with $d_{23}$ = 20 nm. The mode (SPhPs and HMs) wavelength of SiO$_2$ and VO$_2$ are denoted with filled squares in the figure.

To understand the underlying mechanism responsible for the rectification function, we plotted the spectral radiative heat flux in Figs. 3(a) and (b) for TD and GTD, respectively, with $d_{23}$ = 20 nm. The frequency of $SiO_2$ surface phonon polaritons (SPhPs) and $VO_2$ modes are denoted by the filled squares in the figures. Fig. 3(a) shows that for the TD system, the heat flux peaks mainly near the $SiO_2$ SPhPs band, but has almost no response in the spectrum of $VO_2$ modes. There is no significant difference in the heat flux between the forward-biased and reverse-biased scenarios. However, the results are quite different in Fig. 3(b). For the forward-biased scenario, with the adding of graphene, two obvious peaks of spectral heat flux emerge in the spectrum of $VO_2$ modes. Additionally, the original two peaks in the region of $SiO_2$ SPhPs also greatly enhance compared to the TD. At the same time, the heat flux in the reverse-biased scenario tends to be uniform in the spectral range of interest, which is weakly enhanced compared to the TD system. Based on the above, it can be inferred that the improved performance of GTD, is mainly attributed to the interactions between graphene and $SiO_2/VO_2$, which greatly enhaces the heat flux in the forward-biased scenario by forming coupled and stronger modes.

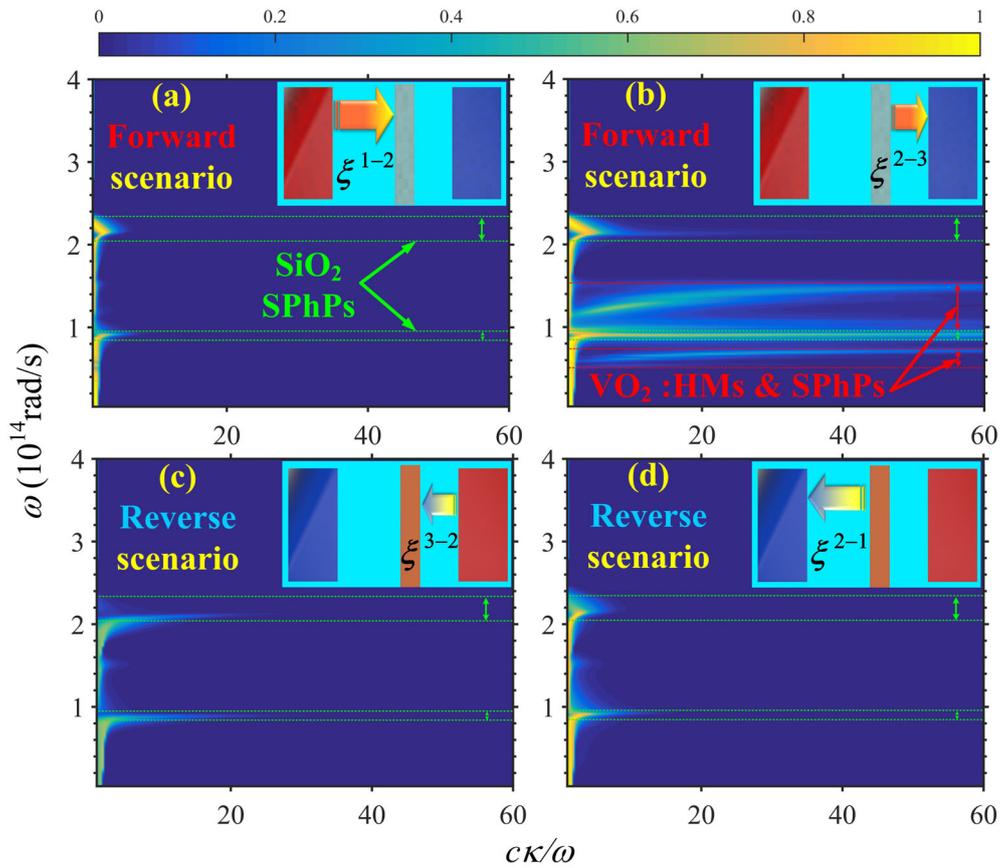

Fig. 4. The p-polarized energy transmission coefficients for thermal diode without graphene (TD) in forward and reverse scenarios.

The transmission probability of photons indirectly reflects the distribution of the heat flow from a microscopic point of view. To gain a deeper insight into the physics involved in the proposed device, we have plotted the p-polarized energy transmission coefficients for TD in the three-body system in Fig. 4. Furthermore, the modes excited by SiO$_2$ and VO$_2$ are indicated in the figures with dashed lines in green and red, respectively. It is known that for SiO$_2$, two frequency bands exist for SPhPs excited at vacuum / SiO$_2$ interface, from $8.67 \times 10^{13}$ rad/s to $9.47 \times 10^{13}$ rad/s and from $2.03 \times 10^{14}$ rad/s to $2.35 \times 10^{14}$ rad/s[61]. As for VO$_2$, it bahaves like a metal above the phase transition, and there are nearly no surface modes near the interface. However, two different kinds of modes are supported by insulating VO$_2$, namely SPhPs and hyperbolic modes (HMs)[23]. SPhPs can be excited at the vacuum / insulating VO$_2$ surface in the frequency regions where the real parts of both ordinary dielectric function $\varepsilon_\parallel$ and extraordinary dielectric function $\varepsilon_\perp$ are negative. When the real parts of $\varepsilon_\parallel$ and $\varepsilon_\parallel$ have different signs, i.e., (Re ($\varepsilon_\parallel$) × Re ($\varepsilon_\parallel$) < 0), the HMs come into play.

According to the theory of the three-body system, the energy exchange in the system depends on the energy transmission coefficients composed of two different parts. For the forward scenario, $\xi^{1\text{-}2}$ represents the Landauer transmission probability responsible for the difference in thermal populations between the hot body and the VO$_2$ film. Nevertheless, the other part, $\xi^{2\text{-}3}$, is more complicated, as it contains the transmission coefficients of the intermediate body $\tau_2$ and the reflection amplitude associated to the couple of bodies 1 and 2 treated as a unique body[33]. The formulas for these coefficients are similar to Eqs. (8)-(9) and are given in Ref [33], and thus, will not be repeated here.

In Figs. 4(a) and (b), we plotted the contours of the two different parts of p-polarized energy transmission coefficients for TD in the forward scenario. It is shown that in Fig. 4(a), the SPhPs of SiO$_2$ dominate the energy transmission coefficients between the hot body and the VO$_2$ slab, but there is no response at the spectrum of VO$_2$ modes. However, for $\xi^{2\text{-}3}$ in Fig. 4(b), which represents the transmission probability of photons between the pair (1, 2) and body 3, besides the SiO$_2$ SPhPs, the SPhPs and HMs of VO$_2$ also emerge. It can be inferred that the emergence of VO$_2$ modes is mainly caused by the coupling between bodies 1 and 3, which are made of the same material. The resonance between the two silica slabs provides an opportunity for VO$_2$ modes to be excited and coupled with them. However, the modes of VO$_2$ are still quite weak, resulting in low heat flux in the spectrum of VO$_2$ modes in Fig. 3(a). In contrast to the forward scenario, the energy transmission coefficients of the reverse scenario in Figs. 4(c) and (d) are both weaker. There are only contours of SiO$_2$ SPhPs, but nearly no response in the VO$_2$ spectrum. The vanishing of VO$_2$ modes is due to the fact that it is metallic this time and can hardly interact with the SiO$_2$ SPhPs. Not only that, the metallic VO$_2$ destroys the coupling between the two silica slabs and hence breaks the channel for photon tunneling. It acts like a block in the three-body system, resulting in lower heat flux compared to the forward scenario. The basic rules of how the phase transition material plays a role in the diode can be preliminarily explained by the above physics.

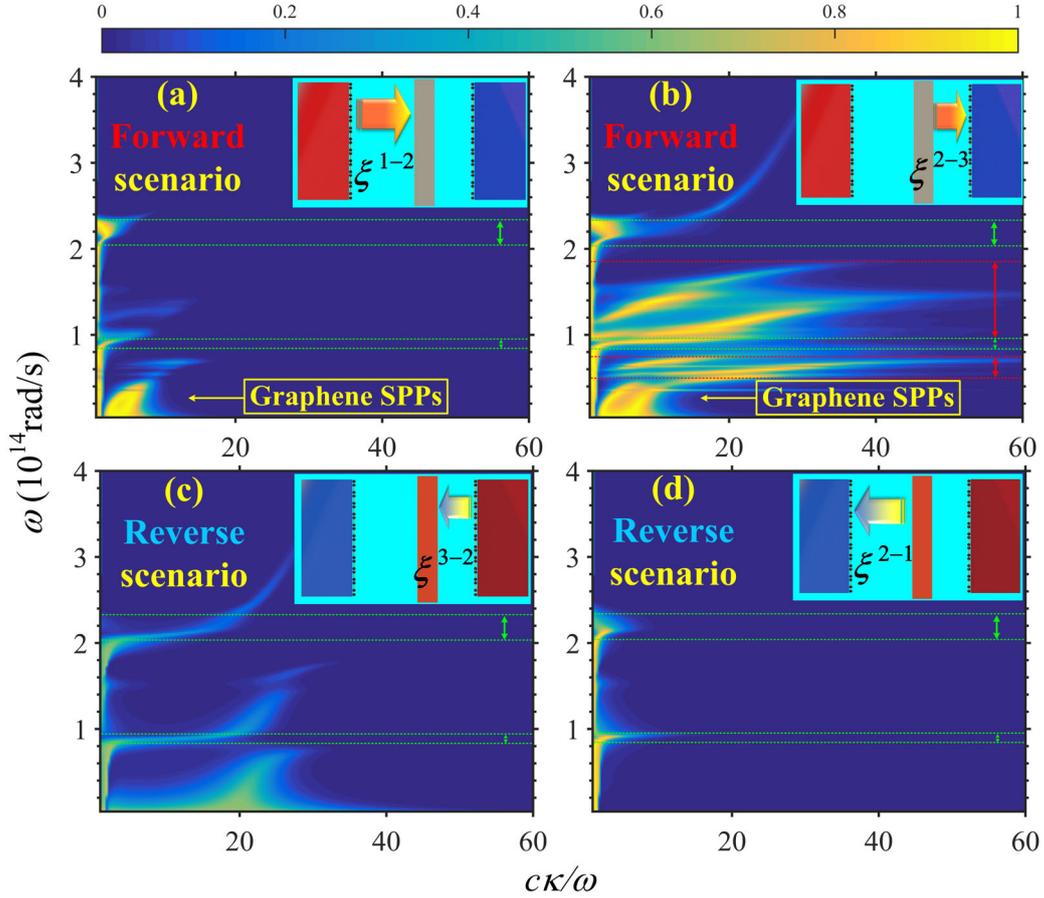

Fig. 5. The p-polarized energy transmission coefficients for GTD (with the chemical potentials of graphene set as 0.9 eV) in forward and reverse scenarios.

To better understand the operating principles of the proposed GTD and the role of graphene, we plot the p-polarized energy transmission coefficients as a function of the chemical potential of graphene, which we set to 0.9 eV in Fig. 5. The inclusion of graphene leads to the emergence of an additional branch in the low-frequency regime for the forward scenario, which is due to the excitation of graphene SPPs. This indicates that the two graphene sheets on either side of the system can couple with each other when the intermediate VO$_2$ slab is in an insulating state. Importantly, the energy transmission coefficients shown in Fig. 5(b) are significantly higher than those in Fig. 4(b) without graphene. The results demonstrate that the modes excited by VO$_2$ and SiO$_2$ are both enhanced and spread to larger wave vectors because they can interact with graphene SPPs to form coupled modes. Moreover, the enhancement of energy transmission coefficients is particularly pronounced for the $VO_2$ modes. However, for the reverse scenario in Fig. 5(d), there are only two branches of modes, which correspond to the SPhPs modes of silica. Furthermore, the coefficients are even weaker in Fig. 5(c) despite the presence of a faint branch of graphene SPPs at low frequencies. The breakdown of coupling between the graphene SPPs can be attributed to the metallic $VO_2$, which prevents the two graphene sheets from interacting with each other. Based on the mechanism discussed above, we infer that the operating principles of the proposed GTD are primarily determined by the joint contribution of graphene and the phase transition material. The insulating $VO_2$ can strongly interact with graphene and form coupled modes in the forward scenario, whereas the metallic $VO_2$ acts like a barrier and inhibits the coupling in the reverse scenario. Compared to the TD, the improved performance of the GTD is undoubtedly due to the inclusion of graphene, which stimulates insulating $VO_2$ to participate more actively in heat transfer by photon tunneling.

**B. Effect of Graphene Properties and Geometric Parameters**

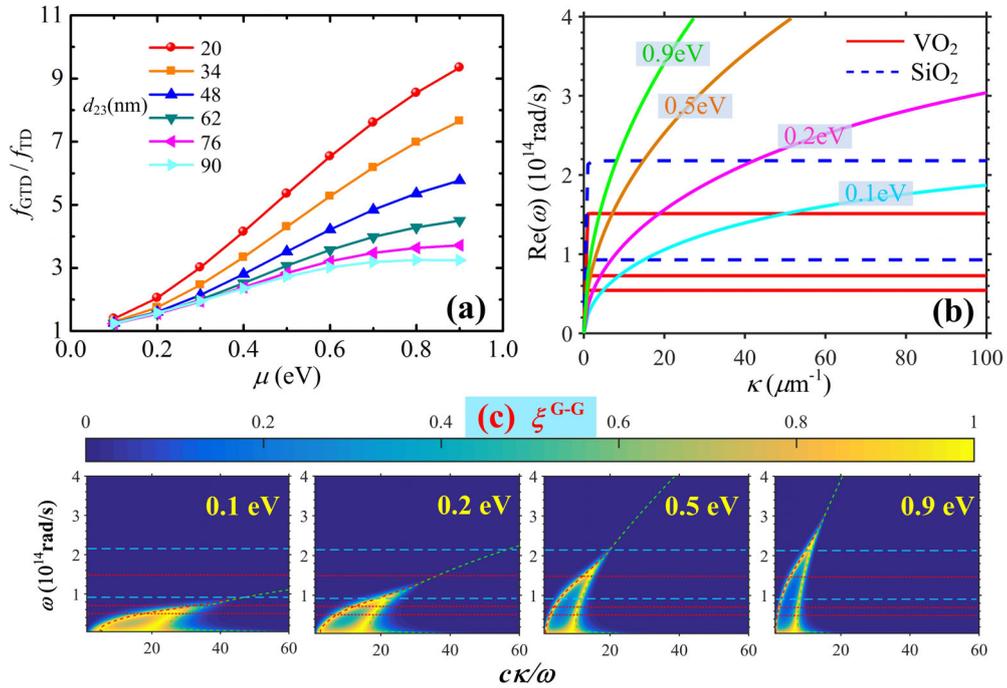

Fig. 6. (a) The improvement of rectification factors caused by graphene with different chemical potentials. (b) Dispersion relations of the modes for $VO_2$, $SiO_2$ and graphene with different chemical potentials. (c) The p-polarized energy transmission coefficients between two suspended graphene sheets with the separation distance

$D+\delta$.

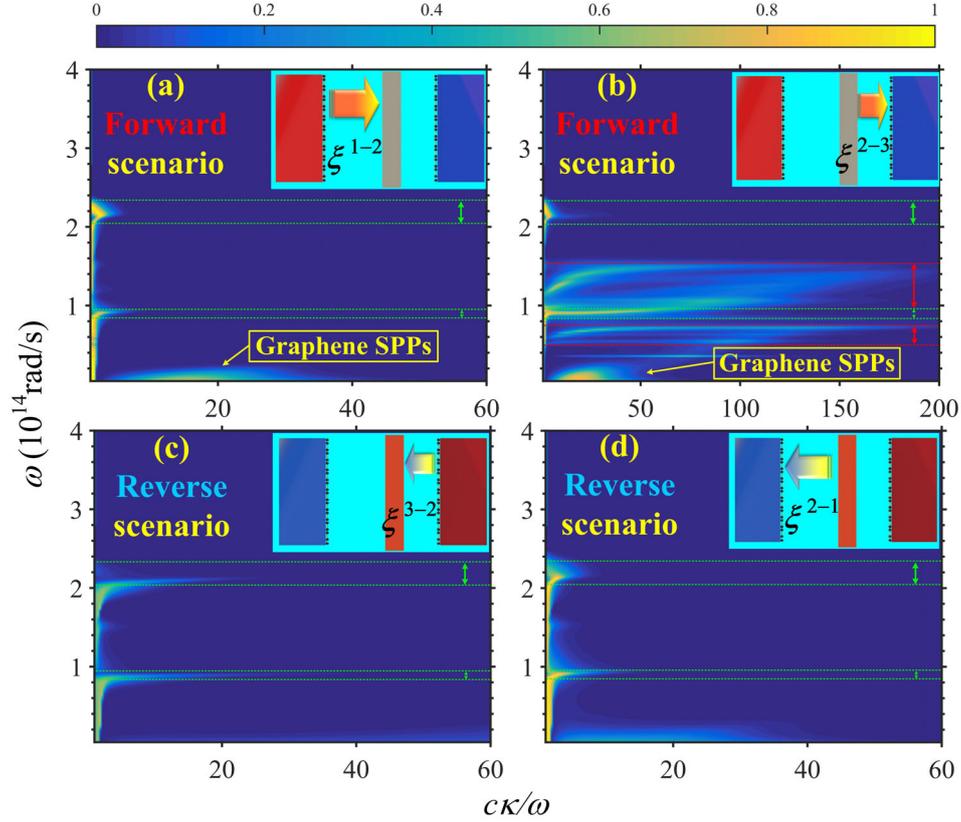

Fig. 7. The p-polarized energy transmission coefficients for GTD (with the chemical potentials of graphene set as 0.1 eV) in forward and reverse scenarios.

The fundamental physical property of graphene is its chemical potential, which can be modulated from 0 eV to 1eV by applying a bias voltage[48]. To investigate the effect of graphene's properties on the performance of the proposed GTD, we plot the enhanced rectification factors $f_{GTD} / f_{TD}$ with respect to the chemical potentials of graphene in Fig. 6(a). The results show that the improvement driven by graphene increases with its chemical potential, regardless of the position of the $VO_2$ slab located between the two terminals. To explore the underlying mechanism responsible for this phenomenon, we plot the dispersion relations in Fig. 6(b) for the modes of $VO_2$, $SiO_2$, and graphene with different chemical potentials. Additionally, the p-polarized energy transmission coefficients between two suspended graphene sheets $\xi^{G-G}$ are also plotted in Fig. 6(c) to avoid the influence of the other materials, with the separation distance $D+\delta$, i.e., the distance between the two graphene sheets in Fig. 1(a). The frequency bands corresponding to the modes of $VO_2$ and $SiO_2$ are also shown in Fig. 6(c) with dashed lines. The low-frequency symmetric and high-frequency antisymmetric branches of two suspended graphene sheets are denoted by the dashed lines in green and red, respectively[66]. The results show that with a low chemical potential of graphene $\mu$ = 0.1 eV, the dispersion relations and the energy transmission coefficients of graphene SPPs both concentrate at low frequencies and intersect with the modes of $VO_2$ or $SiO_2$ only in tiny parts. However, as the chemical potential increases, the graphene SPPs spread to higher frequncies and the intersecting parts become larger. When the chemical potential reaches $\mu$ = 0.9 eV, the branches of graphene SPPs in Fig. 6(c) completely cover the frquency bands of the modes supported by $VO_2$ and $SiO_2$. This means that as the chemical potential increases, graphene can interact with $VO_2$ and $SiO_2$ more intensively and hence form stronger coupled modes, which is the key to the realization of the proposed thermal diode.

To verify the above analysis, we plot the p-polarized energy transmission coefficients for GTD with a chemical potential of graphene set at 0.1 eV in Fig. 7. Compared to the results in Fig. 5 with $\mu$ = 0.9 eV, the values of energy transmission coefficients become much smaller. The graphene SPPs tend to lower frequencies and become weaker. Most notably, the enhancement induced by the coupling between graphene SPPs and $VO_2$ modes observed in Fig. 5(b) has almost vanished, indicating that the most important factor responsible for the performance of the GTD has diminished. These results indicate that coated graphene should be selected with higher chemical potentials to achieve a better performance of the proposed GTD.

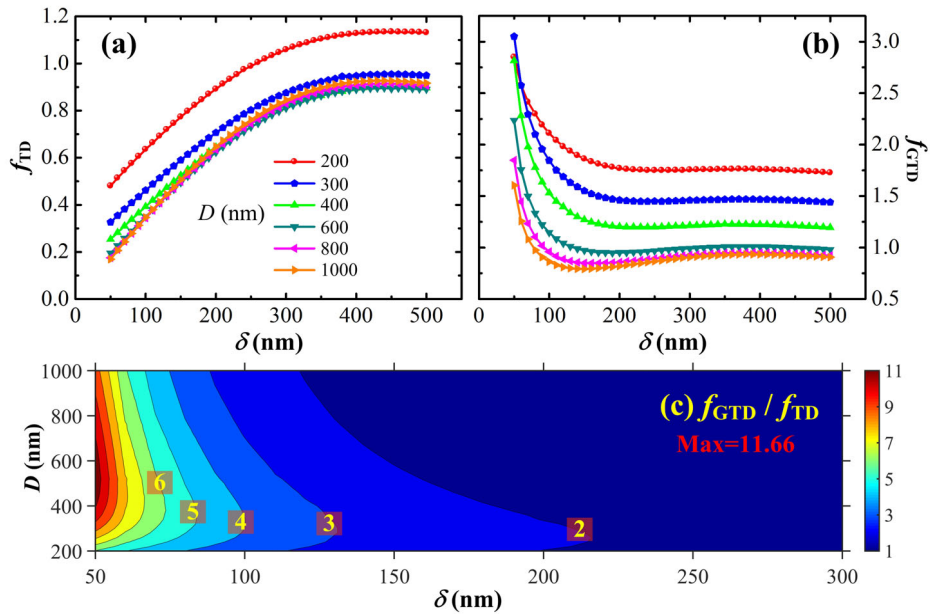

Fig. 8   The rectification factors for different vacuum separations with respect to the thickness of the $VO_2$ membrane, for (a) TD and (b) GTD; (c) The ratio of rectification factors between GTD and TD for different

geometric parameters of the three-body system.

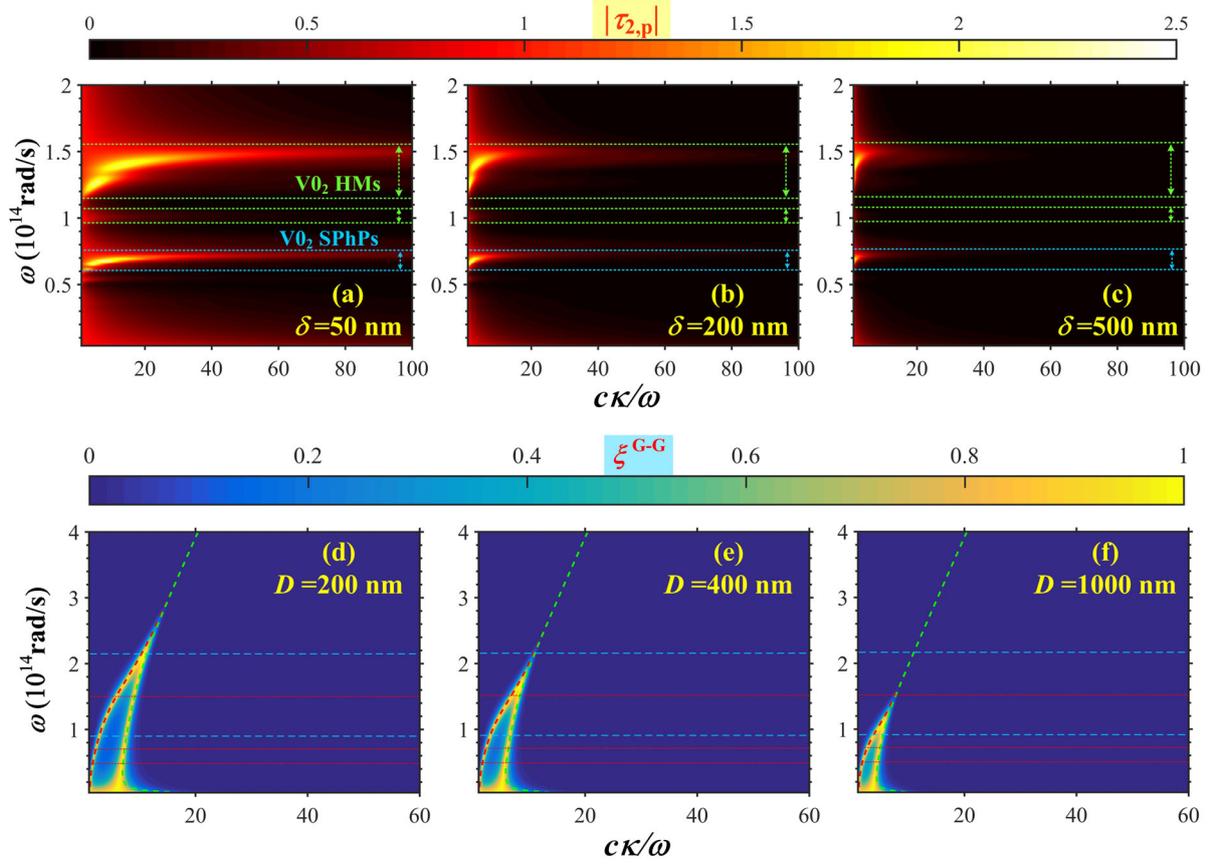

Fig. 9. (a) Transmission coefficients $|\tau_{2,p}|$ of the intermediate VO$_2$ membrane with different thicknesses. (b) The p-polarized energy transmission coefficients between two suspended graphene sheets (with the chemical potentials of graphene set as 0.9 eV) with the separation distance $D+\delta$ ($\delta = 200$ nm).

In addition to the properties of graphene, the geometric parameters of the system also influence the performance of the GTD. The most effective geometric parameters for the three-body system are the thickness of the intermediate body and the total vacuum separation distance between the external two terminals. Figs. 8(a) and (b) show the rectification factors for different vacuum separations with respect to the thickness of the VO$_2$ membrane for TD and GTD, respectively. The results demonstrate that they exhibit significantly different trends with respect to the thickness of the VO$_2$ membrane. The rectification factor for TD ($f_{TD}$) gradually increases as the thickness becomes larger and remains nearly unchanged when the thickness is greater than 400 nm. Conversely, the rectification factor for GTD ($f_{GTD}$) dramatically decreases at samll $\delta$ and then remains nearly unchanged at $\delta >$ 200 nm. However, regardless of whether it is TD or GTD, the performance always improves with a smaller vacuum separation. To demonstrate the improvement caused by the addition of graphene, the ratio of rectification factors between GTD and TD is plotted in Fig. 8(c) for different geometric parameters. The improvement is much more significant with a smaller thickness of the $VO_2$ membrane compared to a thicker intermediate body. The maximum value of the ratio can be as high as 11.66, indicating that the addition of graphene can effectively enhance the performance of the thermal diode. In summary, to obtain a better performance of the proposed GTD, one should choose the geometric parameters with a thinner $VO_2$ membrane and a smaller separation distance between the two terminals.

Here, we thoroughly investigate the phenomenon responsible for the above observations. Figs. 9(a) – (c) show the transmission coefficients of the intermediate $VO_2$ membrane $|\tau_{2,p}|$ for thickness of $\delta$ = 50 nm, 200 nm, and 500 nm, respectively. The results reveal that as the $VO_2$ membrane becomes thicker, the transmission coefficients weaken at the frequency bands accounting for the $VO_2$ HMs and SPhPs. This implies that the two terminals on both sides of the $VO_2$ membrane can hardly interact with each other when the $VO_2$ membrane is too thick. Moreover, for the two coated graphene sheets, which are the key component to realize the function of the proposed GTD, the coupling between them weakens as $|\tau_{2,p}|$ dissipates. Hence, the performance of the GTD degrades with a thicker $VO_2$ membrane. In Figs. 9(d) – (f), we plot the p-polarized energy transmission coefficients between two suspended graphene sheets ($\mu$ = 0.9 eV) with the separation distance $D+\delta$ for $D$ = 200 nm, 400 nm, and 1000 nm, respectively. The results show that as the separation distance increases, the branch of graphene SPPs tends to lower frequencies, which results in weaker interactions between graphene and $VO_2$. This is confirmed in Figs. 9(d) – (f) by the intersections between graphene SPPs and the modes of $VO_2$ indicated by the dashed lines. As is known for graphene SPPs, they are surface modes that can only be excited near the interface and decay rapidly along the normal direction away from the interface[48]. Due to the confirmed fact that the graphene SPPs assist the modes of $VO_2$ to form the reticfication function, the performance of the GTD will undoubtedly degrade when the graphene SPPs weaken.

## IV. Conclusions

In this work, we propose a graphene-assisted radiative thermal diode based on the three-body photon thermal tunneling framework, and the rectification factor can reach 300 % with a 350 nm separation distance between the two terminals of the diode. We find that the performance of the radiative thermal diode is significantly improved by incorporating graphene, with the improvement being more than 11 times. The operating principles of the proposed GTD are attributed to the combined contribution of graphene SPPs and the phase transition material.

The insulating VO$_2$ can strongly interact with graphene to form coupled modes in the forward scenario, while the metallic VO$_2$ acts as a barrier and disrupts the coupling in the reverse scenario. We also discover that the performance of the proposed GTD is enhanced with higher chemical potentials of graphene, thinner VO$_2$ slabs, and smaller vacuum separation distances. Our proposed device indicates that the addition of an intermediate relay body can slow down the performance degradation of the thermal diode, and has overcome the limitation of using phase transition material as terminals. As a result, the fabrication and operation of the thermal diode can be easier in practical applications at the nanoscale. This device may pave the way for the development of thermal-photon-based logical circuits, thermal photon-driven communication technologies, and nanoscale thermal management approaches.

**Declaration of competing interest**

The authors declare that they have no known competing financial interests or personal relationships that could have appeared to influence the work reported in this paper.

**Data availability**

Data will be made available on request.

**Acknowledgements**

The supports of this work by the National Natural Science Foundation of China (No. 52206082), China Postdoctoral Science Foundation (No. 2021TQ0086), the Natural Science Foundation of Heilongjiang Province (No. LH2022E063), Postdoctoral Science Foundation of Heilongjiang Province (No. LBH-Z21013) are gratefully acknowledged.